\documentclass[12pt]{article}
\usepackage{graphicx}

\newcommand\pubnumber{}
\newcommand\pubdate{\today}

\def\nagoya{Kobayashi-Maskawa Institute,\\
Nagoya University, Nagoya, JAPAN}

\def\Title#1{\begin{center} {\Large #1 } \end{center}}
\def\Author#1{\begin{center}{ \sc #1} \end{center}}
\def\Address#1{\begin{center}{ \it #1} \end{center}}

\newcommand\pubblock{\rightline{\begin{tabular}{l} \pubnumber\\
         \pubdate  \end{tabular}}}
\newenvironment{Abstract}{\begin{quotation}  }{\end{quotation}}
\newenvironment{Presented}{\begin{quotation} \begin{center} 
             PRESENTED AT\end{center}\bigskip 
      \begin{center}\begin{large}}{\end{large}\end{center} \end{quotation}}

\def\beq{\begin{equation}}
\def\eeq#1{\label{#1}\end{equation}}
\def\eeqn{\end{equation}}

\def\beqa{\begin{eqnarray}}
\def\eeqa#1{\label{#1}\end{eqnarray}}
\def\eeqan{\end{eqnarray}}

\let\bar=\overbar

\def\Dslash{\not{\hbox{\kern-4pt $D$}}}
\def\dslash{\not{\hbox{\kern-2pt $\del$}}}

\def\msb{{\bar{\ssstyle M \kern -1pt S}}}

\begin{document}
\begin{titlepage}
\pubblock

\vfill
\Title{Inclusive $B$ decays and exclusive radiative decays by Belle}
\vfill
\Author{Yutaro Sato}
\Address{\nagoya}
\vfill
\begin{Abstract}
The $b \rightarrow s \gamma$, $b \rightarrow d \gamma$ and $b \rightarrow s \ell^+ \ell^-$ processes
are allowed at higher order via the electroweak loop or box diagrams in the Standard model.
It is sensitive probe to search for new physics beyond the Standard model
because new particles might enter in the loop.

We present preliminary results of
branching fraction of the $\bar{B} \rightarrow X_s \gamma$,
$CP$ asymmetry in the $\bar{B} \rightarrow X_{s+d} \gamma$,
and the forward-backward asymmetry in the $\bar{B} \rightarrow X_s \ell^+ \ell^-$.
The results are based on a data sample
containing $772 \times 10^6 B\bar{B}$ pairs recorded at the $\Upsilon(4S)$ resonance with the Belle detector at the KEKB $e^+ e^-$ collider.
\end{Abstract}
\vfill
\begin{Presented}
The 8th International Workshop \\
on the CKM Unitarity Triangle (CKM 2014)\\
Vienna, Austria, September 8--12, 2014
\end{Presented}
\vfill
\end{titlepage}
\def\thefootnote{\fnsymbol{footnote}}
\setcounter{footnote}{0}

\section{Introduction}
The $b \rightarrow s \gamma$,
$b \rightarrow d \gamma$ and $b \rightarrow s \ell^+ \ell^-$ processes
are allowed at higher order via the electroweak loop or box diagrams in the Standard model (SM).
It is sensitive probe to search for new physics beyond the SM
because new particles might enter in the loop.
In this report,
we present results of inclusive or semi-inclusive measurements about
electroweak penguin processes. 
Inclusive measurement are preferable to exclusive measurements
because of lower theoretical uncertainties,
although they are experimentally more challenging.

\section{Branching Fraction of the $\bar{B} \rightarrow X_s \gamma$}
For this analysis a ``sum of exclusive'' approach is chosen,
i.e. 
we measure as many exclusive $X_s$ modes as possible
and then sum them up to extrapolate inclusive branching fraction.
We reconstruct the $B$ meson
from a high energy photon and one of the 38 $X_s$ final states.
We require the photon candidate
with energy $1.8$ GeV $< E_{\gamma}^* < 3.4$ GeV in the center-of-mass (CM) frame.

The dominant background comes from $e^+ e^- \rightarrow q \bar{q} (u,d,s,c)$
continuum events,
which is suppressed
using mainly event shape information.
For an effective background rejection,
we employ a neural network
based on the software package
``NeroBayes'' package \cite{NB}.
When $\pi^0$ from the $\rho$ emits a high energy photon
in the $B \rightarrow D^{(*)} \rho^+$ decay,
it looks like the signal.
To veto such backgrounds,
we reconstruct $D$ candidates of the major decay modes
with combinations of particles used in the $X_s$ reconstruction,
and veto events with reconstructed $D$ mass close to the nominal $D$ mass.

\begin{figure}[htb]
\centering
\includegraphics[height=7cm]{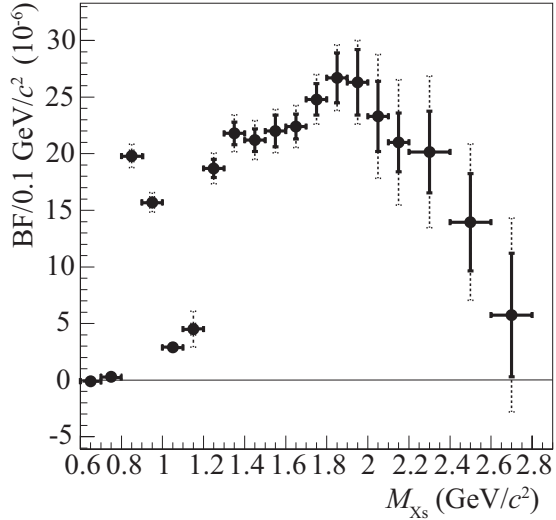}
\caption{Partial branching fraction as a function of $M_{X_s}$.
The error basrs correspond to the statistical (solid)
and the quadratic sum of the statistical and systematic (dashed).}
\label{fig:xsgamma_br}
\end{figure}

The signal yields are extracted by an maximum likelihood fit to the beam-constrained mass, $M_{\rm bc}$.
To minimize the systematic uncertainty from modeling of the $X_s$ mass distribution,
we divide the data into 19 bins of $X_s$ mass in the region
$0.6$ GeV/$c^2 < M_{X_s} < 2.8$ GeV/$c^2$.
Maximum $X_s$ mass corresponds to a minimum photon energy of 1.9 GeV.
Figure~\ref{fig:xsgamma_br} shows the partial branching fraction as a function of $M_{X_s}$.
Total branching fraction in $M_{X_s} < 2.8$ GeV/$c^2$
is obtained from the sum of 19 $M_{X_s}$ bins:
\begin{eqnarray}
{\cal B}(\bar{B} \rightarrow X_s \gamma) &=& (3.51 \pm 0.17 \pm 0.33) \times 10^{-4},
\end{eqnarray}
where the first uncertainty is statistical and the second is systematic.
To compare theoretical prediction, the experimental result is extrapolated
to photon energy in the $B$ rest frame above 1.6 GeV
with extrapolation factor
\cite{Xsgamma_br_extrapolation}:
\begin{eqnarray}
{\cal B}(\bar{B} \rightarrow X_s \gamma) &=& (3.74 \pm 0.18 \pm 0.35) \times 10^{-4},
\end{eqnarray}
which is consistent with SM prediction \cite{Xsgamma_br_prediction}
within $1.3\sigma$
and the most precise result of any sum-of-exclusives approach.

\section{$CP$ asymmetry in the $\bar{B} \rightarrow X_{s+d} \gamma$}
The $CP$ asymmetry (${\cal A}_{\rm CP}$) in the $\bar{B} \rightarrow X_{s+d} \gamma$ decays is defined as:
\begin{eqnarray}
{\cal A}_{\rm CP} = 
\frac{\Gamma(\bar{B} \rightarrow X_{s+d} \gamma) - \Gamma(B \rightarrow X_{\bar{s}+\bar{d}} \gamma)}{\Gamma(\bar{B} \rightarrow X_{s+d} \gamma) + \Gamma(B \rightarrow X_{\bar{s}+\bar{d}} \gamma)}.
\end{eqnarray}
In the SM, ${\cal A}_{\rm CP}(\bar{B} \rightarrow X_{s+d} \gamma)$ is predicted to be zero
with negligible theoretical uncertainty
\cite{Xsgamma_acp_prediction}.
In this analysis, we reconstruct only high energy photon
with an CM energy
$1.4$ GeV $< E_{\gamma}^* < 2.8$ GeV
from signal-$B$ decay
and perform a fully inclusive measurement.
To identify the flavor of signal-$B$ meson, we use charge of a lepton,
which comes from semileptonic decay of opposite $B$ meson.
We require lepton momentum with $1.10$ GeV $< p_{\ell}^* < 2.25$ GeV in the CM frame.

\begin{figure}[htb]
\centering
\includegraphics[height=7cm]{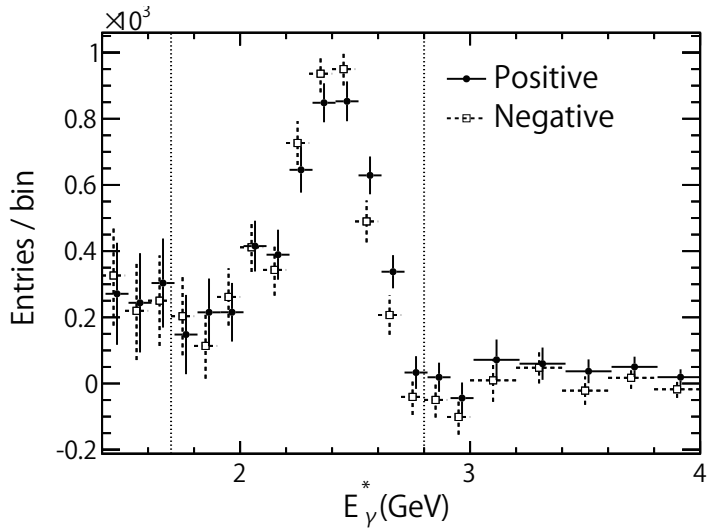}
\caption{Photon energy spectrum in the CM frame after background subtraction.
The positive (circles) and negative (squares) tagged events are shown.
Error bars includes statistical and systematic uncertainties.}
\label{fig:xsgamma_acp}
\end{figure}

The signal is extracted by subtracting
the $B \bar{B}$ and continuum backgrounds.
The continuum contribution is subtracted using the off-resonance data.
Dominant $B\bar{B}$ background comes from the $\pi^0$ and $\eta$ decay,
which is calibrated from Monte Carlo samples
with correction factors in $\pi^0$ and $\eta$ momentum bins.
Figure~\ref{fig:xsgamma_acp} shows the photon energy spectrum in the CM frame after background subtraction.
From this we calculate ${\cal A}_{\rm CP}$,
correcting possible asymmetry from detector and $B \bar{B}$ background
and dilution from $B \bar{B}$ mixing.
The measurement of ${\cal A}_{\rm CP}$ is performed
for photon energy thresholds
between 1.7 GeV and 2.2 GeV.
For the $E_{\gamma}^* > 2.1$ GeV,
we obtain 
\begin{eqnarray}
{\cal A}_{\rm CP}(\bar{B} \rightarrow X_{s+d} \gamma)
=
(2.2 \pm 4.0 \pm 0.8)\%,
\end{eqnarray}
where first uncertainty is statistical
and the second is systematic.
This result is consistent with the SM prediction and the most precise measurement.

\section{Forward-backward Asymmetry in the $B \rightarrow X_s \ell^+ \ell^-$}
The lepton forward-backward asymmetry is defined as
\begin{eqnarray}
  {\cal A}_{\rm FB}(q^2_{\rm min},q^2_{\rm max}) =
   { \int_{q^2_{\rm min}}^{q^2_{\rm max}} dq^2 \int_{-1}^1 d\cos\theta
   \;{\rm sgn}(\cos\theta) {d^2 \Gamma \over dq^2 d\cos\theta} \over
     \int_{q^2_{\rm min}}^{q^2_{\rm max}} dq^2 \int_{-1}^1 d\cos\theta
       {d^2 \Gamma \over dq^2 d\cos\theta} },
\label{eq:afb}
\end{eqnarray}
where $q^2 = M_{\ell^+\ell^-}^2$
and $\theta$ is the angle between the $\ell^+ (\ell^-)$
and the $B$ meson momentum in the $\ell^+ \ell^-$ center-of-mass frame
in $\bar{B}^0$ or $B^-$ ($B^0$ or $B^+$) decays.
For this analysis, a ``sum of exclusive'' approach is chosen.
We reconstruct $B$ meson from dilepton with opposite charge and one of 10 $X_s$ final states, where lepton is a electron or a muon.
The inclusive $B \rightarrow X_s \ell^+ \ell^-$ is extrapolated from the sum of 10 exclusive $X_s$ states,
assuming ${\cal A}_{\rm FB}$
does not depend on the lepton flavor and $X_s$ mass.
To reject a large part of the combinatorial background,
we require $M_{X_s} < 2$ GeV/$c^2$.

The main background comes from random combinations of two semileptonic $B$ or $D$ decays,
which have both large missing energy due to neutrinos,
and displaced origin of leptons from $B$ or $D$ mesons.
Other background originates from continuum events,
which is suppressed using event shape variables.
To efficiently suppress semileptonic and continuum backgrounds,
we employ the NeuroBayes. 

\begin{figure}[htb]
\centering
\includegraphics[height=7cm]{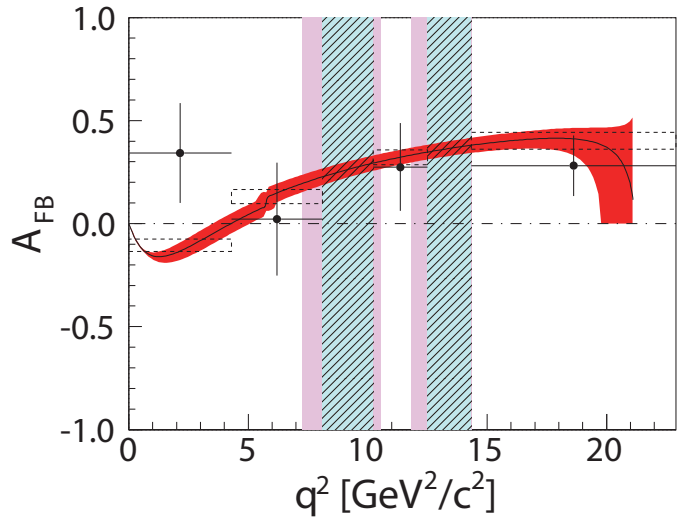}
\caption{
Measured ${\cal A}_{\rm FB}$ as a function of $q^2$.
The curve (black) with the band (red)
and dashed boxes (black) represent the SM prediction \cite{Xsll_afb_prediction1, Xsll_afb_prediction2}
while filled circles with error bars show the fit results.
The $J/\psi$ and $\psi$(2S) veto regions are shown as teal hatched regions. For the electron channel,
the pink shaded regions are added to the veto regions dueto the large bremsstrahlung effect.
}
\label{fig:xsll_afb}
\end{figure}

To examine the $q^2$ dependence of ${\cal A}_{\rm FB}$,
we divide the data into 4 bins of measured $q^2$.
In order to extract ${\cal A}_{\rm FB}$,
maximum likelihood fit to four $M_{\rm bc}$ distributions (positive/negative $\cos\theta$ for electron/muon channel) is simultaneously performed for each $q^2$ bin.
The signal reconstruction efficiency dependence to $q^2$ and $\cos\theta$ is taken into account.
Figure~\ref{fig:xsll_afb} shows the measured ${\cal A}_{\rm FB}$ as a function of $q^2$.
For $q^2 > 10.2$ GeV$^2/c^2$,
${\cal A}_{\rm FB} < 0$ is excluded at the 2.3$\sigma$ level.
For $q^2 < 4.3$ GeV$^2/c^2$,
the result is within 1.8$\sigma$
of the SM expectation.
This result is the first measurement of 
${\cal A}_{\rm FB}$ in inclusive $B \rightarrow X_s \ell^+ \ell^-$.

\section{Conclusion}
We report precise measurement of branching fraction of the $\bar{B} \rightarrow X_s \gamma$
and 
$CP$ asymmetry in the $\bar{B} \rightarrow X_{s+d} \gamma$,
and
first measurement of the forward-backward asymmetry
in the $\bar{B} \rightarrow X_s \ell^+ \ell^-$.
The results based on the large data sample
recorded by the Belle detector at the KEKB $e^+ e^-$ collider.
All results are compatible with the SM expectation.
Analysis about
$CP$ asymmetry in the $\bar{B} \rightarrow X_{s+d} \gamma$,
and
the forward-backward asymmetry
in the $\bar{B} \rightarrow X_s \ell^+ \ell^-$
are limited by statistics
and thus will be measured more precisely at Belle II.
More precise measurement of branching fraction of the $\bar{B} \rightarrow X_s \gamma$ will be performed
with different approach
in which the other $B$ meson is fully reconstructed at Belle II. 



\begin{thebibliography}{99}


\bibitem{NB}
M.~Feindt and U.~Kerzel,
Nucl. Instrum. Methods Phys. Res.,
Sect. A {\bf 559}, 190 (2006).

\bibitem{Xsgamma_br_extrapolation}
O.~Buchmuller and H.~Flacher,
Phys. Rev. {\bf D73}, 073008 (2006).

\bibitem{Xsgamma_br_prediction}
M.~Misiak {\it et al.}
Phys. Rev. Lett. {\bf 98}, 022002 (2007).

\bibitem{Xsgamma_acp_prediction}
M.~Benzke, S.J.~Lee, M.~Neubert and G.~Paz,
Phys. Rev. Lett. {\bf 106}, 141801 (2011).

\bibitem{Xsll_afb_prediction1}
A.~Ali, E.~Lunghi, C.~Greub, and G.~Hiller,
Phys. Rev. {\bf D66}, 034002 (2002).

\bibitem{Xsll_afb_prediction2}
S.~Fukae, C.S.~Kim, T.~Morozumi, and T.~Yoshikawa
Phys. Rev. {\bf D59}, 074013 (1999).

\end{thebibliography}
\end{document}